# Power-law distributions and fluctuation-dissipation relation in the stochastic dynamics of two-variable Langevin equations

Du Jiulin

Department of Physics, School of Science, Tianjin University, Tianjin 300072, China
E-mail: jiulindu@yahoo.com.cn

**Abstract**: We show that the general two-variable Langevin equations with inhomogeneous noise and friction can generate many different forms of power-law distributions. By solving the corresponding stationary Fokker-Planck equation, we can obtain a condition under which these power-law distributions are accurately created in a system away from equilibrium. This condition is an energy-dependent relation between the diffusion coefficient and the friction coefficient and thus it provides a fluctuation-dissipation relation for nonequilibrium systems with power-law distributions. Further, we study the specific forms of the Fokker-Planck equation that correctly leads to such power-law distributions, and then present a possible generalization of Klein-Kramers equation and Smoluchowski equation to a complex system, whose stationary-state solutions are exactly a Tsallis distribution.

# 1. Introduction

In some complex systems away from equilibrium, there are a variety of non-exponential or power-law distributions that have been frequently observed and studied. These distributions have been noted prevalently in, for example, glasses [1,2], disordered media [3-5], folding of proteins [6], single-molecule conformational dynamics [7,8], trapped ion reactions [9], chemical kinetics, and biological and ecological population dynamics [10,11], reaction-diffusion processes [12], chemical reactions [13], combustion processes [14], gene expressions [15], cell reproductions [16], complex cellular networks [17], small organic molecules [18], astrophysical and space plasmas [19] etc. The typical forms of such power-law distributions which have attracted great attention include the $\kappa$-distributions or the generalized Lorentzian distributions in the solar wind and space plasmas [19-24], the $q$-distributions in complex systems within nonextensive statistical mechanics [25, 26], and those $\alpha$-distributions noted in physics, chemistry and elsewhere like $P(E) \sim E^{-\alpha}$ with an index $\alpha > 0$ [9,12,13, 18, 20, 27]. These power-law distributions may lead to processes different from those in the realm governed by the traditional Boltzmann-Gibbs statistics with Maxwell-Boltzmann (M-B) distribution. Simultaneously, a class of statistical mechanical theories studying the power-law distributions in complex systems has been constructed, for instance, by generalizing Boltzmann entropy to Tsallis entropy [25], by generalizing Gibbsian theory [28] to a system away from thermal equilibrium, and so forth. A correct and deep understanding of the details of stochastically dynamical origins underlying these power-law distributions has become very important endeavors for understanding the nature in many different processes of physical, chemical, biological, technical and their inter-disciplinary fields. In these aspects, some interesting theoretical and experimental research works have been reported in recent and previous diverse studies [20, 25, 26, 9, 29-45]. However, the dynamical origins and causes of power-law distributions remain unclear. An open problem, which is being investigated intensively, is the conditions under which those power-law distributions are produced and the physical mechanism that leads to such distributions in complex systems away from equilibrium.

Theoretically, the power-law distributions have been studied so far mainly through single-variable Langevin equations and/or the associated Fokker-Planck (F-P) equations [29-35, 39, 45] (of course, there are also some Boltzmann equation and master equation approaches). This might usually be because it is difficult to solve a general multi-variable F-P equation. In fact, not much is known in general about the long time stationary-state solution of an arbitrary F-P equation. As early as 1985, through introducing the photon-induced Coulomb-field fluctuations in a superthermal radiation field plasma, the novel power-law $\kappa$-distribution observed in space plasmas was produced by employing the velocity-space F-P equation [20]. Later, with the introduction and development of nonextensive statistical mechanics, initiated by Tsallis in 1988 [25], one has tried to explain the dynamical origins of $q$-distributions [30]. In a single-variable Ito-Langevin equation, a possible condition was analyzed under which the corresponding F-P equation might have the stationary-state solution of $q$-distribution [31]. A velocity-space linear Langevin equation with fluctuations in temperature was constructed so that it could yield a dynamical reason for a $q$-distribution [34]. By introducing multiplicative fluctuations in a momentum-space linear Langevin equation, one could lead to a simple $q$-distribution [39]. Most recently, one-variable Langevin equations both for position-space and momentum-space as well as the corresponding F-P equations have been discussed, and the general conditions have been analyzed under which power-law distributions may be produced [46]. Another way to study the stochastic dynamics of power-law distributions is the $q$-generalized Boltzmann equation [36-38]. By constructing a generalized $q$-equilibrium condition for self-gravitating long-range interaction systems, one could generate the $q$-distributions in several of physical situations [42], among which the power-law distribution function for a nonextensive system was derived. Using this distribution function, one has excellently modeled the dark matter haloes observed on spherical galaxies with regard to their nonequilibrium stationary state [47]. In the meantime, several experimental studies of power-law distributions have been carried out, such as a test of $q$-distributions by solar sound speeds in helioseismological measurements [40], a demonstration of momentum distribution in dissipative optical lattices [41], and a detection of anomalous diffusion in driven-dissipative dusty plasma [43,44].

It is well known that the stochastic dynamics to generate a M-B distribution for a thermal equilibrium system can be well explained by the two-variable Langevin equations for position and momentum, (x, p), which models the motion of a Brownian particle moving in a potential field and in the medium with a constant friction coefficient and a white noise. And the diffusion constant is related with the friction constant by the well-known fluctuation-dissipation relation (FDR). Questions are naturally raised for a complex system away from equilibrium. Can there be general two-variable Langevin equations that have a stochastic dynamics to generate power-law distributions? Can there also be a FDR under which a power-law distribution is created? And can we find a stochastic differential equation that has a stationary-state solution of power-law distribution? For these questions, in this work, we try to study the underlying stochastic dynamics of power-law distributions by giving general two-variable Langevin equations and the corresponding F-P equation. In Section 2, by introducing inhomogeneous noise and friction into the Langevin equations and then solving the corresponding stationary F-P equation, we derive a stationary-state solution of power-law distribution. The generalized FDR, an energy-dependent relation between the diffusion coefficient and the friction coefficient, is determined, under which many different forms of power-law distributions can be created in a system away from equilibrium. In Section 3, we study a possible generalization of Klein-Kramers equation and Smoluchowski equation to a nonequilibrium system with power-law distribution. Finally in Section 4, we give our conclusions and discussions.

**2. The underlying stochastic dynamics of power-law distributions**

Let us consider the general two-variable Langevin equations for position and momentum (x, p), which is known to model the Brownian motion of a particle moving in a potential field $V(x)$,

$$\frac{dx}{dt} = \frac{p}{m}, \qquad \frac{dp}{dt} = -\frac{dV(x)}{dx} - \gamma\, p + \eta(x,p,t), \tag{1}$$

where $m$ is the particle's mass, and $\gamma$ is the friction coefficient. Usually in normal conditions, the friction coefficient is regarded as a constant. But when the Brownian particle moves in an inhomogeneous complex medium, it may be considered as a function of the variables $x$ and $p$, i.e. $\gamma = \gamma(x, p)$. For a complex system, the noise may be considered as inhomogeneous in $(x, p)$ and so it is also a function of the variables $x$ and $p$, i.e. $\eta(x,p,t)$ is a multiplicative (space/velocity dependent) noise [48]. Having given the Langevin equations, one is often faced with the dilemma how to interpret them with the noises [49]. Different interpretations lead to different drift terms in the corresponding F-P equation. The examples are the F-P equations in the conventional Stratonovich and Ito rules. For a complex system which is driven away from equilibrium through the interactions with the external field and the noise, the F-P equation for noise-averaged distribution may apply even more to a description of the macroscopic distribution. As usual, one can assume that the noise is Gaussian, with zero average and the delta-correlated in time $t$, such that it satisfies

$$\langle \eta(x, p, t) \rangle = 0, \quad \langle \eta(x, p, t) \eta(x, p, t') \rangle = 2D(x, p) \delta(t - t'). \tag{2}$$

The correlation strength of multiplicative noise (i.e. diffusion coefficient) $D(x, p)$ is a function of the variables $x$ and $p$.

Using the given two-variable Langevin equations with the assumption Eq.(2) made for the noise, we can now follow Zwanzig's rule [50] to write the corresponding F-P equation for the noise-averaged distribution. The formation of F-P equation in Zwanzig's rule appears similar to that in the Stratonovich rule [49, 51] but without the "spurious drift". If $\rho(x, p, t)$ is the probability distribution function of variables $(x, p)$ at time $t$, corresponding to the Langevin equations (1) and Eq.(2), the F-P equation for the noise-averaged distribution can be written as

$$\frac{\partial \rho}{\partial t} = -\frac{p}{m}\frac{\partial \rho}{\partial x} + \frac{\partial}{\partial p}\left(\frac{dV(x)}{dx} + \gamma(x, p)\, p\right)\rho + \frac{\partial}{\partial p}\left(D(x, p)\frac{\partial \rho}{\partial p}\right). \tag{3}$$

Its stationary-state solution, $\rho_s(x, p)$, satisfies the stationary F-P equation,

$$-\frac{p}{m}\frac{\partial \rho_s}{\partial x}+\frac{\partial}{\partial p}\left(\frac{dV}{dx}+\gamma\, p\right)\rho_s+\frac{\partial}{\partial p}\left(D\frac{\partial \rho_s}{\partial p}\right)=0. \tag{4}$$

(Here and hereafter, for economization of space we leave out the independent variables of functions in the equations). Now we look back on the case where, when the friction coefficient and the diffusion coefficient are both constants and they are related with each other by the standard FDR, the solution of this equation is the M-B distribution, which is known a function of the energy. Now in the case that the friction coefficient and diffusion coefficient are both a function of the variables $x$ and $p$, as a macroscopic distribution of a complex system, we may seek a stationary-state solution of Eq.(4) in the form of $\rho_s(x,p)=\rho_s(E(x,p))$ with the particle's energy $E(x, p) \equiv V(x)+p^2/2m$. Namely, we assume that the stationary distribution is a function of the energy. Such a consideration seems reasonable because, according to current knowledge, the stationary distribution is often a function of energy, such as the M-B distribution, the Tsallis distribution, and the $\kappa$-distribution observed. Thus, from Eq.(4) one can derive that

$$\frac{\partial}{\partial p}\left(\gamma\, p\rho_s+\frac{p}{m}D\frac{d\rho_s}{dE}\right)=0. \tag{5}$$

After integrating this equation for $p$, one has

$$p\left(\gamma\, \rho_s+\frac{D}{m}\frac{d\rho_s}{dE}\right)=C(x), \tag{6}$$

where $C(x)$ is an "integral constant". Because $C(x)$ is independent of $p$, setting $p=0$ on the left hand side of Eq.(6), one can determine $C(x)=0$. And then, the stationary-state solution is solved exactly,

$$\rho_s(E)=Z^{-1}\exp\left(-m\int\frac{\gamma}{D}dE\right), \tag{7}$$

where $Z$ is a normalization constant.

Now we discuss the possible forms of the integrand in Eq.(7) in order to produce a suitable power-law distribution. It is well known that, when a system reaches its thermal equilibrium state, there exists a standard FDR, i.e. $D/m\gamma=\beta^{-1}$ with $\beta=1/kT$, and Eq.(7) is the famous M-B distribution. When a system is away from equilibrium, this simple FDR will often be used in many physical situations (e.g. in chemical reaction systems). However, the fact is that this might be correct only under normal conditions, while for the anomalous diffusion in complex systems this is not the case. Some examples of anomalous diffusion imply that the diffusion coefficient might not only depend on kinetic energy (the square of momentum or velocity) [20, 30, 32, 35, 39] but also depend on potential energy [45, 51-55]. Recently, an experimental study of anomalous diffusion in driven-dissipative dusty plasma was carried out, presenting a power-law $q$-distribution and a strong insight into dependence of the diffusion coefficient on the interaction potential [43, 44]. Accordingly, as a reasonable assumption for a complex system, we may let ($D/m\gamma$) be a function of the energy generally when the system is away from equilibrium, i.e. $D/m\gamma = f(E)$, while $f(E) = \beta^{-1}$ when the system reaches a thermal equilibrium. Such a $f(E)$ should be a positive function because there is no negative diffusion coefficient $D$ and there is no negative friction coefficient $\gamma$. In principle, it is possible to determine this function by experimental measures for real nonequilibrium complex systems. We now discuss this function and the resulting stationary-state distributions in the following two cases (a) and (b).

(a). $f(E)$ is a general function of $E$. Theoretically, if $f(E)$ is a continuous and differentiable function at least in the surrounding of $E=0$, then it can be expressed as the Maclaurin series, i.e., $f(E) = f(0) + f'(0)E + \cdots + f^{(n)}(0)E^n/n! + \cdots$, where $f'(0) = \left[\partial f(E)/\partial E\right]_{E=0}$, …, and $f^{(n)}(0) = \left[\partial^n f(E)/\partial E^n\right]_{E=0}$ with $n$ positive integer. There is a situation that one can determine the distribution function exactly. For a low energy system, if the energy $E$ is small so that the terms of $n \geq 2$ in the series can be neglected, then the function becomes $f(E) \approx f(0) + f'(0)E = \beta^{-1} + f'(0)E$, where the first term in the series has been denoted by $f(0) = \beta^{-1}$. Hence we find that

$$D = m\gamma\beta^{-1}(1 - \kappa\beta E), \tag{8}$$

where we have introduced a parameter $\kappa \equiv -f'(0)$. When $\kappa=0$, $f(E)$ is a constant and Eq.(8) is the standard FDR in the equilibrium. Obviously, the parameter $\kappa \neq 0$ measures a distance away from thermal equilibrium. Eq.(8) provides a relation between the friction coefficient and the diffusion coefficient for a system away from equilibrium, and therefore it is a generalization of FDR. Substituting this generalized FDR Eq.(8) into Eq.(7), one can show that the stationary-state solution becomes the power-law $\kappa$-distribution, i.e.

$$\rho_s(E) = Z^{-1}(1 - \kappa\beta E)_+^{1/\kappa}, \tag{9}$$

where the normalization constant is $Z = \iint dxdp \left(1 - \kappa\beta E\right)_+^{1/\kappa}$, and $(y)_+ = y$ for $y > 0$ and zero otherwise. Actually, even for $\kappa > 0$, it is not necessary to prescribe the meaning of $(y)_+$ because $E$ is small here and $f(E)$ does not change its positive nature by neglecting the higher order terms. Eq.(9) has exactly the form of $\kappa$-distribution observed in the solar wind and space plasmas [19-24, 28] (or Tsallis distribution in nonextensive statistical mechanics [25]) and in the limit $\kappa \to 0$ it recovers the M-B distribution. It is clear that the generalized FDR Eq.(8) is a condition under which the power-law $\kappa$-distribution can be created from the stochastic dynamics of the two-variable Langevin equations. The standard FDR can be recovered by Eq.(8) in the case of $\kappa = 0$, and therefore Eq.(8) is the FDR for nonequilibrium systems with the power-law $\kappa$-distribution.

(b). $f(E)$ is only a linear function of $E$. In this case, we can let $f(E) = \beta^{-1} + \mu E$, where the correlation coefficient should be $\mu \geq 0$ so as to ensure that $f(E)$ is a positive function for all values of the energy. And then we find

$$D = m\gamma\beta^{-1}(1 + \mu\beta E), \tag{10}$$

and the stationary-state solution Eq.(7) becomes a power-law $\mu$-distribution, i.e.

$$\rho_s(E) = Z^{-1}(1 + \mu\beta E)_+^{-1/\mu}, \tag{11}$$

where $Z = \iint dxdp \left(1 + \mu\beta E\right)_+^{-1/\mu}$. Eq.(11) is also equivalent to the $\kappa$-distribution observed in the solar wind and space plasmas (or Tsallis distribution in nonextensive statistical mechanics) and in the limit $\mu \to 0$ it recovers the M-B distribution. The parameter $\mu \neq 0$ also measures a distance away from thermal equilibrium. Eq.(10) is a condition under which the power-law $\mu$-distribution is created from the stochastic dynamics of the two-variable Langevin equations, and thus it is also a generalization of FDR to the nonequilibrium systems with power-law $\mu$-distribution.

If the energy $E$ is high enough (for the high energy system with linear positive correlation of $f(E)$ to $E$) so that $\beta^{-1}$ can be neglected as compared with the term $\mu E$, then the function is $f(E) \approx \mu E$, and hence we have

$$D = m\gamma \, \mu E . \tag{12}$$

In this case, the stationary-state solution Eq.(7) becomes a power-law $\alpha$-distribution,

$$\rho_s(E) = A\left(\frac{E}{E_0}\right)^{-\alpha}, \tag{13}$$

with a parameter $\alpha$ defined by $\alpha \equiv 1/\mu$, where $A$ is the normalization constant, and $E_0$ is an appointed reference energy so as to ensure the quantity in the bracket to be dimensionless. The power-law $\alpha$-distributions have been frequently observed and noted in some systems away from equilibrium, such as the space plasma at a high energy [20], the solar flares [27], the chemical processes and reaction-diffusion processes [12,13,18], the trapped ion collisions with heavy neutrals [9], and elsewhere. Eq.(12) is also a FDR in the systems with power-law $\alpha$-distribution, which provides a condition under which the power-law $\alpha$-distribution is created.

The physical mechanism that generates the power-law $\kappa$-distribution by Eqs.(8) and (9) can well explain a momentum-space diffusion coefficient, $D(p)=D\left(1+\frac{C}{D}p^2\right)$ which led to a Tsallis velocity distribution from a one-variable linear Langevin equation and the corresponding F-P equation (see Eq.(20) in Re.[39] ). This diffusion coefficient can be correctly described by the generalized FDR Eq.(8) in the case when one sets the potential $V(x)=0$, the parameter $\kappa = -(2mC/D\beta)$, while the equilibrium diffusion coefficient $D = m\gamma\beta^{-1}$.

The physical mechanism Eq.(8) of generating the power-law $\kappa$-distribution also well works for explaining the power-law distribution theory given by Hasegawa et al in the study of super-thermal radiation field plasma [20]. The theory gave rise to the famed $\kappa$-distribution observed in space plasmas through introducing the photon-induced Coulomb-field fluctuations in a one-variable F-P equation for the Coulomb field. The longitudinal velocity-space diffusion coefficient of a test particle was derived by

$$D_\|(\upsilon) = D_\|^{eq(e)}\left(1 + \frac{D_\|^{NL(e)}}{D_\|^{eq(e)}}\right) = D_\|^{eq(e)}\left(1 + \frac{k_D^2 |r_0|^2}{3|\varepsilon(\omega_0,0)|^2 \ln \Lambda} \cdot \frac{\upsilon^2}{\upsilon_{Te}^2}\right), \quad (14)$$

where $D_\|^{eq(e)}$ is the diffusion coefficient due to the equilibrium field, $D_\|^{NL(e)}$ is the diffusion coefficient due to the photon-induced longitudinal field, $k_D$ is the Debye wave number, $r_0$ is the direction-averaged excursion distance of electrons in the photon electric field, $\varepsilon$ is the longitudinal plasma dielectric function, $\ln \Lambda$ is the Coulomb logarithm, and $\upsilon_{Te}$ is the electron thermal velocity. It is easy to see that Eq.(14) is exactly equal to the generalized FDR Eq.(8) in the case without the potential $V(x)$ and if one sets our $\kappa$–parameter (the parameter $\kappa$ here is precisely the inverse of that one in Re.[20]) in the form

$$\kappa = -\frac{2k_D^2 |r_0|^2}{3|\varepsilon(\omega_0,0)|^2 \ln \Lambda}, \quad (15)$$

while the equilibrium diffusion coefficient is $D_\|^{eq(e)} = \gamma\beta^{-1}$ in terms of the standard FDR, where the test particle mass is $m=1$ and then the mean square thermal velocity is $\upsilon_{Te}^2 = \beta^{-1}$.

In the same way, the physical mechanism which generates the power-law $\mu$-distribution by Eqs.(10) and (11) is also well works for the above two examples.

Theoretically, a non-exponential stationary-state solution exists and is available from the stationary F-P equation whenever

$$m\gamma / D = f^{-1}(E) \quad (16)$$

is an integrabel and positive function of the energy in Eq.(7). Thus, Eq.(16) is a more general FDR for complex systems with non-M-B distribution. We conclude that the stationary distribution of the system is a M-B distribution if $f(E)$ is a constant, it is a power-law distribution if $f(E)$ is a linear function of the energy, and it is a general non-exponential distribution if $f(E)$ is a nonlinear function of the energy, the results of which as given here only depend on two assumptions, that the inhomogeneous noise satisfies Eq.(2) and that the stationary distribution is a function of the energy. If insert Eq.(16) back in Eq.(7), we readily obtain a FDR which depends on the stationary distribution $\rho_s(E)$, i.e.

$$\frac{m\gamma}{D} = -\frac{d\ln\rho_s(E)}{dE}. \tag{17}$$

Such an expression confirms a guess of nonequilibrium generalized FDR which indeed depends on an invariant distribution (see Eq.(3.9) in Ref.[56]), and it seems equivalent to that FDR if the invariant distribution is a function of the energy. An observable dependence of FDR-temperature was seen recently when the invariant distribution (non-uniform stationary distribution) is only a function of the energy [57].

## 3. A generalization of Klein-Kramers equation and Smoluchowski equation

Now we study the FDR Eq.(8) in order to write a specific form of the F-P equation that correctly leads to the power-law $\kappa$-distribution (in the same way, we can also study the FDR Eq.(10) to write a specific form of the F-P equation that correctly leads to the power-law $\mu$-distribution). For this purpose, as combined with the power-law $\kappa$-distribution Eq.(9), Eq.(8) can be equivalently written as $D = m\gamma\beta^{-1}(Z\rho_s)^\kappa$, and then one has

$$\frac{\partial}{\partial p}\left(D\frac{\partial \rho_s}{\partial p}\right) = Z^\kappa \cdot \frac{1}{\kappa+1}m\gamma\beta^{-1}\frac{\partial^2}{\partial p^2}\rho_s^{\kappa+1}. \tag{18}$$

For simplicity and convenience of writing, the coefficients in the front of the term on the right hand side of Eq.(18) can be identified with one coefficient $D_\kappa$ by setting the relation $Z^{\kappa} = (\kappa+1)D_{\kappa}\beta(m\gamma)^{-1}$. Substituting Eq. (18) into Eq.(4), the stationary F-P equation can be written by

$$-\frac{p}{m}\frac{\partial \rho_s}{\partial x} + \frac{\partial}{\partial p}\left(\frac{dV}{dx} + \gamma p\right)\rho_s + D_{\kappa}\frac{\partial^2}{\partial p^2}\rho_s^{\kappa+1} = 0. \tag{19}$$

As one expected, the solution of this equation is surely the power-law $\kappa$-distribution Eq.(9). If one lets $\kappa=0$, Eq.(19) becomes the usual stationary Klein-Kramers equation [48], whose solution is a M-B distribution. In this sense, one can present a possible generalization of Klein-Kramers equation to the nonequilibrium systems with the power-law $\kappa$-distribution,

$$\frac{\partial \rho}{\partial t} = -\frac{p}{m}\frac{\partial \rho}{\partial x} + \frac{\partial}{\partial p}\left(\frac{dV}{dx} + \gamma p\right)\rho + D_{\kappa}\frac{\partial^2 \rho^{\kappa+1}}{\partial p^2}, \tag{20}$$

whose stationary-state solution is the $\kappa$-distribution [19-24,28] (or Tsallis distribution in nonextensive statistical mechanics[25]). This equation is exactly the Klein-Kramers equation [48, 58] if one takes the parameter $\kappa=0$. Some single-variable F-P equations whose forms are similar to Eq.(20) have been the object of diverse studies [29, 45, 52, 53, 59]. It is remarkable that the time-dependent solutions of these equations are still a function of the energy (i.e. potential energy or kinetic energy) if the initial condition has a delta-function on the variable [45]. Nevertheless, it remains to be proved whether the time-dependent solution of Eq.(20) is only a function of the energy.

Let us consider the physical situation where the Brownian particle moves in a strong friction medium [48, 50, 58]. In this case, the coordinate undergoes a creeping motion, and the derivative on momentum may be eliminated approximately from the Langevin equation Eq.(1), leading to the following stochastic dynamical equations only for the coordinate,

$$\frac{dx}{dt} = -(m\gamma)^{-1}\frac{dV(x)}{dx} + (m\gamma)^{-1}\eta(x,t), \tag{21}$$

with the noise to satisfy

$$\langle\eta(x,t)\rangle = 0, \quad \langle\eta(x,t)\eta(x,t')\rangle = 2D(x)\delta(t-t'). \tag{22}$$

The diffusion coefficient is $D(x)$. The time evolution of the corresponding reduced probability distribution, $\rho(x,t) = \int_{-\infty}^{\infty} dp \rho(x,p,t)$, is governed by the corresponding one-variable F-P equation,

$$\frac{\partial \rho}{\partial t} = (m\gamma)^{-1} \frac{\partial}{\partial x}\left(\frac{dV}{dx}\rho\right) + (m\gamma)^{-1} \frac{\partial}{\partial x}\left(D \frac{\partial \rho}{\partial x}\right). \tag{23}$$

Its stationary-state solution, $\rho_s(x)$, is easily given by

$$\rho_s(x) \sim \exp\left(-\int \frac{1}{D(x)} dV(x)\right). \tag{24}$$

It is clear that when the system reaches the equilibrium, $D(x)$ is a constant, $D=\beta^{-1}$, and Eq.(24) is a Boltzmann distribution. When the system is away from equilibrium, one may consider $D(x)$ as a function of the potential energy $V(x)$, i.e. $D(x) = f(V(x))$. In the same way as previously led to Eq.(8) for the power-law $\kappa$-distribution (the way that leads to Eq.(10) for the power-law $\mu$-distribution can also apply here similarly), after expanding the function $f(V)$ as a Maclaurin series, one can show that if the diffusion coefficient satisfies

$$D(x) = \beta^{-1}[1 - \kappa\beta V(x)] \tag{25}$$

then the stationary-state solution Eq.(24) becomes the power-law $\kappa$-distribution,

$$\rho_s(x) = Z_\kappa^{-1}\left[1 - \kappa\beta V(x)\right]_+^{1/\kappa}, \tag{26}$$

where $Z_\kappa$ is a normalization constant: $Z_\kappa = \int dx \left[1 - \kappa\beta V(x)\right]_+^{1/\kappa}$ with a parameter $\kappa = -f'(0)$. The parameter $\kappa \neq 0$ also measures the distance away from thermal equilibrium. Eq.(26) becomes a Boltzmann distribution if the parameter is $\kappa \to 0$. Eq.(25) is also a FDR and it is the condition under which the power-law $\kappa$-distribution can be generated from the stochastic dynamics in Eqs.(21) and (22). In order to write the specific form of the F-P equation that correctly leads to the power-law $\kappa$-distribution Eq.(26), combining Eq.(25) with Eq.(26) one writes $D = \beta^{-1}(Z_\kappa \rho_s)^\kappa$. The stationary F-P equation is

$$(m\gamma)^{-1} \frac{\partial}{\partial x}\left(\frac{dV}{dx}\rho_s\right) + (\kappa+1)^{-1}(m\gamma\beta)^{-1} Z_\kappa^\kappa \frac{\partial^2 \rho_s^{\kappa+1}}{\partial x^2} = 0, \tag{27}$$

whose solution is surely the power-law $\kappa$-distribution (or Tsallis distribution). Eq.(27) is exactly the stationary Smoluchowski equation [48, 50, 58] if one sets $\kappa=0$. For simplicity, if let the coefficients in the front of the second term on the left hand side of Eq.(27) be associated with a constant $D_\kappa$ by $Z_\kappa^\kappa = m\gamma\beta(\kappa+1)D_\kappa$, one obtains a generalization of Smoluchowski equation to the system with the power-law $\kappa$-distribution,

$$\frac{\partial \rho}{\partial t} = (m\gamma)^{-1} \frac{\partial}{\partial x}\left(\frac{dV}{dx}\rho\right) + D_\kappa \frac{\partial^2 \rho^{\kappa+1}}{\partial x^2}, \qquad (28)$$

whose stationary-state solution is exactly Tsallis distribution. If one sets $\kappa=0$, Eq.(28) recovers the Smoluchowski equation. A thermostatistics of interacting particles in an over damped medium is investigated recently through a single-variable nonlinear F-P equation [45], where the stationary-state solution and the time-dependent solution are both a power-law distribution and they are both a function of the energy. Eq.(28) is precisely the same as this nonlinear F-P equation if the potential function chooses $V(x)=\frac{1}{2}ax^2$. Therefore, the generalized Smoluchowski equation Eq.(28) is a more general stochastic differential equation which has the stationary-state / time-dependent solution of power-law distribution.

## 4. Conclusions and discussions

In conclusions, from a general two-variable Langevin equations and the corresponding Fokker-Planck equation, we have studied the stochastic dynamical origins of power-law distributions in a complex system and have given a generalized FDR for a nonequilibrium system with power-law distributions. This FDR is an energy-dependent relation between the inhomogeneous diffusion coefficient and the inhomogeneous friction coefficient, exhibiting a response of the Brownian particle to the interactions with its environment. This FDR is the condition under which many different forms of power-law distributions can accurately be created from the stochastic dynamics in the two-variable Langevin equations. These power-law distributions includes the famous $\kappa$-distribution observed in space plasmas, Tsallis $q$-distribution in nonextensive statistical mechanics and those $\alpha$-distributions noted in physics, chemistry and elsewhere like $P(E) \sim E^{-\alpha}$ with an index $\alpha >0$, or others. We showed that the power-law $\kappa$–distribution appears in a nonequilibrium system only if the FDR function $D/m\gamma = f(E)$ is a linear function of the energy $E$, in which the correlation constant $\kappa \neq 0$ measures a distance away from thermal equilibrium, the power-law $\alpha$–distribution appears in a high energy system only if the FDR function is $f(E) \approx E/\alpha$, and a general non-exponential distribution appears in a complex system only if the FDR function $f(E)$ is a nonlinear function. These results presented a possible stochastic dynamical mechanism for explaining such power-law distributions of different forms in complex systems.

We further studied specific forms of the Fokker-Planck equations that correctly lead to the stationary-state solution of power-law distribution. By using the generalized FDR for the power-law distribution, we obtained a possible generalization of Klein-Kramers equation and Smoluchowski equation to a complex sytem, whose stationary-state solutions are exactly a Tsallis distribution. It is expected that these equations may find their potential applications, leading an insight into the research of reaction rate theory [58, 60] for nonequilibrium complex systems with power-law distributions [46].

At the end of this paper, we give some additional comments. In the light of recent understanding [61], what distinguishes an equilibrium state from a nonequilibrium stationary state is not the form of the stationary distribution, but the presence or the absence of a detailed balance (i.e. so-called probability currents $K_i^j = 0$ or $K_i^j \neq 0$) in the stationary state in the master equation. When making a comparison between the stationary master equation and the stationary F-P equation, one cannot prove yet that Eq.(6) is equal to the detailed balance. There is an example that the velocity-space power-law distribution, generated from the stochastic dynamics of a single-variable linear Langevin equation and the corresponding stationary F-P equation [46], does represent a nonequilibrium stationary state [37, 38]. Nevertheless, because there was no consideration of the nonextensivity of the energy, the usual power-law $q$-distribution for the energy was found just to be an isothermal distribution [42, 62, 63], and therefore it might not be a correct description for a nonequilibrium stationary state of long-range interacting systems. The detailed balance in a master equation and its relation to the stationary state in an F-P equation is an important issue worthy of further study. If Eq.(6) was equal to such a detailed balance, the power-law $\kappa$–distribution (or $\mu$–distribution) would be considered as *an equilibrium stationary state*, and so it would be an isothermal distribution, and the parameter $\kappa \neq 0$ (or $\mu \neq 0$) would become a measure of the distance away from M-B distribution.

**Acknowledgements**

I would like to thank the National Natural Science Foundation of China under Grant No.11175128 for financial support.